\documentclass[12pt]{article}

\usepackage{amssymb}
\usepackage{amsmath}
\usepackage{graphicx}

\def\be{\begin{eqnarray}}
\def\ee{\end{eqnarray}}
\def\eps{\varepsilon}
\def\Li{\mathop{\rm Li}\nolimits}

\begin{document}

\title{Weakly nonadditive Polychronakos statistics}
\author{Andrij Rovenchak}

\maketitle

\abstract{
A two-parametric fractional statistics is proposed, which can be used to model a weakly-interacting Bose-system. It is shown that the parameters of the introduced weakly nonadditive Poly\-chro\-nakos statistics can be linked to effects of interactions as well as finite-size corrections. The calculations of the specific heat and condensate fraction of the model system corresponding to harmonically trapped Rb-87 atoms are made. The behavior of the specific heat of three-dimensional isotropic harmonic oscillators with respect to the values of the statistics parameters is studied in the temperature domain including the BEC-like phase transition point.
}

\section{Introduction}
In recent decades, a number of modifications of the conventional Bose--Einstein and Fermi--Dirac statistics have been proposed. Quantum-mechanical generalizations include anyons \cite{Wilczek:1982,Mashkevich_etal:2004,Ponomarenko&Averin:2010} as well as so called $q$-deformed algebras \cite{Biedenharn:1989,Lavagno&NarayanaSwamy:2010}. Approaches rooted more in the statistical mechanics are represented by \cite{Haldane:1991,Wu:1994,Beciu:2009}. A special subbranch is the nonextensive statistical mechanics based on the Tsallis entropy \cite{Tsallis:1988} and its generalizations \cite{Tsallis_etal:1998,Bashkirov:2006,Olemskoi_etal:2010,Livadiotis&McComas:2013}.

Methods involving fractional statistics concepts proved to be suc\-cess\-ful in the studies of the fractional quantum Hall effect, high-temperature superconductivity \cite{Canright&Johnson:1994}, interacting systems in low dimensions \cite{Batchelor_etal:2006,Anghel:2012}, cold atomic gases \cite{Honer_etal:2012}, in the analysis of nuclear matter \cite{Anghel_etal:2012} and even in models of dark matter \cite{Ebadi_etal:2013}. Nonextensive generalizations for Bose--Einstein and Fermi--Dirac statistics were also developed \cite{Buyukkilic&Demirhan:1993, Buyukkilic_etal:1995,Chen_etal:2002,Aragao-Rego_etal:2003}.
{Note a long-existing terminological confusion between nonextensivity and nonadditivity, which was discussed in detail in \cite{Tsallis:2009}. When the former is sporadically referred to in this work, the nonadditive nature of entropy and arising in this context Tsallis $q$-exponentials are generally meant. The terms \textit{nonextensive statistics} and \textit{nonadditive statistics}, however, continue to be used interchangeably in the scientific literature, cf. \cite{Venkatesan&Plastino:2010}.
}

The idea of this paper is to suggest a fractional-statistical model with parameters being linked to interactions and finite-size effects. To be more specific, a bosonic system is considered and thus the modifications of statistics proceed from a reference Bose-distribution. This primary attention to weakly-interacting Bose-systems is reasoned by an ongoing interest in this issue. It can be demonstrated by recent studies of trapped two-dimensional systems \cite{Hadzibabic_etal:2008}, finite systems \cite{Wang&Ma:2010,Wang_etal:2011,Daily_etal:2012}, bosonic mixtures \cite{van_Schaeybroeck:2013}, and some other approaches to analyze the influence of interactions \cite{Chakrabarti_etal:2012,Sofianos_etal:2013,Watabe&Ohashi:2013}.

The proposed model is based on the Polychronakos statistics \cite{Polychronakos:1996,Mirza&Mohammadzadeh:2010} with the Tsallis $q$-exponential standing instead of the conventional one in the expressions for occupation numbers. 
The functional form of the distribution function is thus introduced phenomenologically with the following physical motivation.
While the nonextensivity can be explained by long-range interactions \cite{Abe&Okamoto:2001} and the Poly\-chro\-na\-kos statistics parameter is related to the number-of-states counting, one should not expect that influences of interactions and finite-size effects can be easily attributed to separate modifications of statistics. For instance, the very Polychronakos statistics parameter with a small imaginary part allows modeling a weak dissipative branch of the excitation spectrum of a Bose-system \cite{Rovenchak:2013b,Rovenchak:2014}.
Phenomenological introduction of the Tsallis statistics is known in various aspects \cite{Lavagno&Pignato:2013,BarisBagci&Oikonomou:2013,Wong&Wilk:2013}
as well.

The paper is organized as follows. The statistics is defined and calculations are outlined in Sec.~\ref{sec:def}. Series expansions of occupation numbers as well as energy with respect to small parameters are present\-ed in Sec.~\ref{sec:series}. Equations for the statistics parameters of the three-dimensional (3D) system of isotropic harmonic oscillators are obtained in Sec.~\ref{sec:3Dho}. Analysis of the critical temperature of a weakly-interacting finite Bose-system and respective calculations of the specific heat in Sec.~\ref{sec:Tc} are followed by calculations for model systems obeying the weakly nonadditive Polychronakos statistics in Sec.~\ref{sec:CN}. Conclusions are given in Sec.~\ref{sec:conclusions}.

\section{Statistics definition}\label{sec:def}
Define the occupation number of the $j$th level of a system with the elementary excitation spectrum $\eps_j$ as
\be\label{eq:ni-def}
n_j = \frac{1}{z^{-1} e_q^{\eps_j/T}-\alpha},
\ee
where $T$ is the temperature and $z$ is the fugacity. The Tsallis $q$-exponential is given by
\be
e_q^{x} = \Big[1+(1-q)x\Big]^{1/(1-q)}
\qquad\textrm{for \ $1+(1-q)x>0$}.
\ee
Properties of this and other related functions are well described in Refs.~\cite{Chen_etal:2002,Yamano:2002,Niven&Suyari:2009}

Further in this work small deviations from the Bose-distribution are considered, so the parameters $q$ and $\alpha$ are represented in the form:
\be\label{eq:qa}
q = 1-b;\qquad \alpha = 1+a
\ee
with $a$ and $b$ being small corrections.

Since the $q$-exponential originating from the nonextensive/nonadditive statistical mechanics are used, the statistics with the occupation numbers defined by (\ref{eq:ni-def})--(\ref{eq:qa}) will be hereafter referred to as the \textit{weakly nonadditive Polychronakos statistics} (WNAPS).

While it might seem more natural to introduce the exponential deformation as 
$e_q^{(\eps_j - \mu)/T}$ instead of $z^{-1} e_q^{\eps_j/T}$ using the chemical potential $\mu$, cf. \cite{Buyukkilic&Demirhan:1993, Buyukkilic_etal:1995}, expressions with fugacity $z$ appear to be mathematically simpler for further analysis. In the weakly nonadditive limit $q\to1$ 
\be
e_q^{(\eps_j-\mu)/T} = 
e_q^{\eps_j/T} e_q^{-\mu/T} \left( 1+ (1-q)\frac{\eps_j\mu}{T^2}\right)
\ee
as no simple factorization of the Tsallis $q$-exponentials exists \cite{Yamano:2002}. The fugacity introduced in Eq. (1) can thus be approximately related to the chemical potential as follows:
\be
z^{-1} \simeq e_q^{-\mu/T} 
\left( 1+ (1-q)\frac{\langle\eps\rangle\mu}{T^2}\right),
\ee
where the $j$-dependence in the right-hand side is suppressed by substituting $\eps_j$ with energy of the reference system (cf. below) per particle $\langle\eps\rangle = E/N$.

Let the reference Bose-system is an ideal gas with spectrum $\eps_j$ and degeneracy $g_j$; the fugacity $z_{\rm B}=e^{\mu_{\rm B}/T}$, where $\mu_{\rm B}$ is the chemical potential, is defined by:
\be
N = \sum_j \frac{g_j}{z_{\rm B}^{-1}e^{\eps_j/T}-1},
\ee 
which is understood in the thermodynamic limit. The condition defining the thermodynamic limit itself depends on the system under consideration and will be specified later.

Calculation of thermodynamic functions is made by a simple pro\-ce\-dure. First, the fugacity is defined as a function of $T$ and $N$ from
\be\label{eq:N=sum}
N = \sum_j g_j n_j
\ee
and then it is inserted into the definition of energy
\be
E = \sum_j \eps_j g_j n_j,
\ee
from which the heat capacity is calculated as the temperature derivative:
\be
C = \frac{dE}{dT}.
\ee

\section{Series expansions}\label{sec:series}
Since small deviations from the reference Bose-system are considered, the fugacities also must be expanded around $z_{\rm B}$. Let 
\be
z = z_{\rm B} + \Delta z_1
\ee
for the weakly nonadditive statistics and
\be
z = z_{\rm B} + \Delta z
\ee
for a weakly-interacting finite Bose-system.

In the approximation linear with respect to small corrections the occupation numbers in the WNAPS read:
\be\label{eq:nj-WNPS}
n_j &=& \frac{1}{(z_{\rm B}+\Delta z_1)^{-1} e_{1-b}^{\eps_j/T}-(1+a)} \\ 
&=&\nonumber 
\frac{1}{z_{\rm B}^{-1}e^{\eps_j/T}-1} +
a\frac{1}{\left[z_{\rm B}^{-1}e^{\eps_j/T}-1\right]^2} +
\frac{b\eps_j^2}{2T^2}
\frac{z_{\rm B}^{-1}e^{\eps_j/T}}{\left[z_{\rm B}^{-1}e^{\eps_j/T}-1\right]^2} \\
&&{}+
\frac{\Delta z_1}{z_{\rm B}}
\frac{z_{\rm B}^{-1}e^{\eps_j/T}}{\left[z_{\rm B}^{-1}e^{\eps_j/T}-1\right]^2}. 
\nonumber
\ee
This can be compared to the occupation number of the interacting Bose-system:
\be\label{eq:nj-Bose}
n_j &=& \frac{1}{(z_{\rm B}+\Delta z)^{-1} e^{(\eps_j+\Delta\eps_j)/T}-1} \\ 
&=&\nonumber 
\frac{1}{z_{\rm B}^{-1}e^{\eps_j/T}-1} -
\frac{\Delta \eps_j}{T}
\frac{z_{\rm B}^{-1}e^{\eps_j/T}}{\left[z_{\rm B}^{-1}e^{\eps_j/T}-1\right]^2} 
+
\frac{\Delta z}{z_{\rm B}}
\frac{z_{\rm B}^{-1}e^{\eps_j/T}}{\left[z_{\rm B}^{-1}e^{\eps_j/T}-1\right]^2}.
\ee

Comparing the summands with $b$ in Eq.~(\ref{eq:nj-WNPS}) and with $\Delta\eps_j$ in Eq.~(\ref{eq:nj-Bose}) one can suggest this is  the nonadditivity parameter $b$ that is (chiefly) responsible for effective accounting of the interaction.

The above expansions can be written in a ``macroscopic'' form using
\be\label{eq:N=sum_Bose}
N = \sum_j g_j n_j = 
\sum_j \frac{g_j}{z_{\rm B}^{-1}e^{\eps_j/T}-1}
\ee
with an auxiliary notation
\be
Q = \sum_j \frac{g_j}{\left[z_{\rm B}^{-1}e^{\eps_j/T}-1\right]^2}
\ee
as follows:
\be
N = N + aQ + \frac{\Delta z_1}{z_{\rm B}} (N+Q) +
\frac{b}{2T^2} \sum_j {g_j\eps_j^2}
\frac{z_{\rm B}^{-1}e^{\eps_j/T}}{\left[z_{\rm B}^{-1}e^{\eps_j/T}-1\right]^2} 
\ee
and
\be
N = N + \frac{\Delta z}{z_{\rm B}} (N+Q) -
\frac{1}{T} \sum_j {g_j\Delta\eps_j}
\frac{z_{\rm B}^{-1}e^{\eps_j/T}}{\left[z_{\rm B}^{-1}e^{\eps_j/T}-1\right]^2} 
\ee

On the other hand, for the energy of a weakly-interacting Bose-system one has
\be
E &=& \sum_j (\eps_j + \Delta\eps_j) g_j n_j \\
&=& E_{\rm B} + \frac{\Delta z}{z_{\rm B}} (E_{\rm B}+D_{\rm B}) +
\sum_j g_j \Delta\eps_j 
\frac{z_{\rm B}^{-1}e^{\eps_j/T}(1-\eps_j/T)-1}{\left[z_{\rm B}^{-1}e^{\eps_j/T}-1\right]^2},
\nonumber
\ee
where
\be\label{eq:E=sum_Bose}
&&E_{\rm B} = \sum_j \eps_j g_j n_j = 
\sum_j \frac{g_j\eps_j}{z_{\rm B}^{-1}e^{\eps_j/T}-1},\\
&&D_{\rm B} = 
\sum_j \frac{g_j\eps_j}{\left[z_{\rm B}^{-1}e^{\eps_j/T}-1\right]^2}.
\ee
The energy of the WNAPS system is given by
\be
E &=& \sum_j \eps_j g_j n_j\\
&=& E_{\rm B} + a D_{\rm B} +\frac{\Delta z_1}{z_{\rm B}} (E_{\rm B}+D_{\rm B}) + \frac{b}{2T^2}
\sum_j g_j \eps_j^3 
\frac{z_{\rm B}^{-1}e^{\eps_j/T}}{\left[z_{\rm B}^{-1}e^{\eps_j/T}-1\right]^2}.
 \nonumber
\ee

In order to link the parameters $a,b$ to the quantities, which char\-ac\-ter\-ize the weakly interacting finite system (namely, the spectrum correction $\Delta \eps_j$ and the fugacity correction $\Delta z$), one can use the fol\-low\-ing set of equations:

\be \label{3eqs}
\textrm{(i)}&&
aQ + \frac{\Delta z_1}{z_{\rm B}} (N+Q) +
\frac{b}{2} \sum_j g_j\left(\frac{\eps_j}{T}\right)^2
\frac{z_{\rm B}^{-1}e^{\eps_j/T}}{\left[z_{\rm B}^{-1}e^{\eps_j/T}-1\right]^2}=0\nonumber\\
\nonumber\\
\textrm{(ii)}&&
\frac{\Delta z}{z_{\rm B}} (N+Q) -
\frac{1}{T} \sum_j {g_j\Delta\eps_j}
\frac{z_{\rm B}^{-1}e^{\eps_j/T}}{\left[z_{\rm B}^{-1}e^{\eps_j/T}-1\right]^2} 
=0
\nonumber\\
\\
\textrm{(iii)}&&
a D_{\rm B} +\frac{\Delta z_1}{z_{\rm B}} (E_{\rm B}+D_{\rm B}) + \frac{b}{2}
\sum_j g_j\eps_j \left(\frac{\eps_j}{T}\right)^2 
\frac{z_{\rm B}^{-1}e^{\eps_j/T}}{\left[z_{\rm B}^{-1}e^{\eps_j/T}-1\right]^2}\nonumber\\
&&\qquad=
\frac{\Delta z}{z_{\rm B}} (E_{\rm B}+D_{\rm B}) +
\sum_j g_j \Delta\eps_j 
\frac{z_{\rm B}^{-1}e^{\eps_j/T}(1-\eps_j/T)-1}{\left[z_{\rm B}^{-1}e^{\eps_j/T}-1\right]^2}
\nonumber
\ee

Eq.~(ii) just allows expressing the $\Delta z$ correcting directly via $\Delta\eps_j$ in the linear approximation. A third equations is thus required as the WNAPS correction to fugacity $\Delta z_1$ is in fact the third parameter.

Before proceeding to the calculations of thermodynamic functions of the WNAPS system, it is worth to estimate the values of $a,b$ for some model or real physical systems.

\section{3D harmonic oscillators}\label{sec:3Dho}
Further in the work, the calculations are performed for isotropic three-dimen\-sion\-al (3D) harmonic oscillators with frequency $\omega$. Such a model describes a system of particles trapped to the isotropic harmonic potential. For convenience, the summation is substituted with integra\-tion according to the following rule:
\be
\sum_j g_j \ldots = \int_0^\infty d\eps\,g(\eps)\ldots,
\ee
where the density of states is
\be
g(\eps) = \frac{1}{(\hbar\omega)^3}\frac{\eps^2}{2}.
\ee
Since $g(0)=0$, the contribution of the ground state $j=0$ must be written explicitly for temperatures corresponding to the Bose--Einstein condensation (BEC) phase.

Eq.~(\ref{eq:N=sum_Bose}) becomes
\be\label{eq:N=int_Bose}
N = n_0 + \frac{1}{(\hbar\omega)^3} \int_0^\infty d\eps\,
\frac{\eps^2/2}{z_{\rm B}^{-1}e^{\eps/T}-1} = 
n_0 + \left(\frac{T}{\hbar\omega}\right)^3 \Li_3 z_{\rm B},
\ee
where the polylogarithm function
\be
\Li_s z = \sum_{k=1}^\infty \frac{z^k}{k^s}.
\ee
Energy (\ref{eq:E=sum_Bose}) equals
\be
E_{\rm B} = 
\hbar\omega \left(\frac{T}{\hbar\omega}\right)^4 \Li_4 z_{\rm B}.
\ee

The condition defining the thermodynamic limit of a 3D harmonic oscillator system reads \cite{Damle_etal:1996}:
\be
\omega N^{1/3} = {\rm const}.
\ee

The correction to the spectrum from the interaction in the case of a delta-like interatomic potential $\Phi({\bf r}) = \lambda \delta({\bf r})$, where $\lambda=4\pi\hbar^2a_s/m$ is a coupling constant and $m$ is the mass of an atom, can be majorized by the following expression (cf. \cite{Rovenchak:2007}):
\be
\Delta\eps_j = \hbar\omega N \frac{\gamma}{j+1}
\ee
with
\be
\gamma = \frac{4}{\sqrt{2\pi}}\,\frac{a_s}{a_{\rm ho}}.
\ee
In the above equations, $a_s$ is the $s$-wave scattering length and $a_{\rm ho} = \sqrt{\hbar/m\omega}$ in the harmonic oscillator length.

Performing integrations in (\ref{3eqs}) one can reduce the equations to
\be
&&a\left(\Li_2 z_{\rm B} - \Li_3 z_{\rm B}\right) + 
\frac{\Delta z_1}{z_{\rm B}} \Li_2 z_{\rm B} +
6 b\Li_4 z_{\rm B} = 0,\nonumber\\
\\
&&a\left(\Li_3 z_{\rm B} - \Li_4 z_{\rm B}\right) + 
\frac{\Delta z_1}{z_{\rm B}} \Li_3 z_{\rm B} +
10 b\Li_5 z_{\rm B} \nonumber\\
&&\qquad= 
\frac{\Delta z}{z_{\rm B}} \Li_3 z_{\rm B} +
\int_0^\infty d\xi\; \xi^2\,\frac{\Delta\eps(\xi)}{T}\,
\frac{z_{\rm B}^{-1}e^{\xi}(1-\xi)-1}{\left[z_{\rm B}^{-1}e^{\xi}-1\right]^2}
\nonumber
\ee
with
\be
\frac{\Delta z}{z_{\rm B}} = \frac{1}{\Li_2 z_{\rm B}}
\int_0^\infty d\xi\, \xi^2\,\frac{\Delta\eps(\xi)}{T}
\,
\frac{z_{\rm B}^{-1}e^{\xi}}{\left[z_{\rm B}^{-1}e^{\xi}-1\right]^2}.
\ee
This yields
\be\label{eq:aA+bB=XY}
a A(z_{\rm B},T) + b B(z_{\rm B},T) = 
\left(\frac{\hbar\omega}{T}\right)^2 N\frac{\gamma}{2}
\Big[X(z_{\rm B},T)+ Y(z_{\rm B},T)\Big],
\ee
where
\be
&&A(z_{\rm B},T)=
\frac{\Li_3 z_{\rm B}}{\Li_2 z_{\rm B}} - 
\frac{\Li_4 z_{\rm B}}{\Li_3 z_{\rm B}},
\\
&&B(z_{\rm B},T)=
10\frac{\Li_5 z_{\rm B}}{\Li_3 z_{\rm B}}-
6\frac{\Li_4 z_{\rm B}}{\Li_2 z_{\rm B}},
\\
&&X(z_{\rm B},T) = 
\frac{1}{\Li_2 z_{\rm B}}
\int_0^\infty d\xi\, \frac{\xi^2}{\xi+\hbar\omega/T}
\,
\frac{z_{\rm B}^{-1}e^{\xi}}{\left[z_{\rm B}^{-1}e^{\xi}-1\right]^2},
\\
&&Y(z_{\rm B},T) = \frac{1}{\Li_3 z_{\rm B}}
\int_0^\infty d\xi\, 
\frac{\xi^2}{\xi+\hbar\omega/T}
\,
\frac{z_{\rm B}^{-1}e^{\xi}(1-\xi)-1}{\left[z_{\rm B}^{-1} e^{\xi}-1\right]^2}.
\ee

The parameters $a,b$ appear thus to be temperature-dependent.
Though, the coefficient functions in (\ref{eq:aA+bB=XY}) are smooth enough, so for calculations in a specific temperature domain the value of $T$ can be fixed as shown in the next Section.

On the other hand, it is straightforward to show that in the limit of $T\to\infty$ the fugacity tends to zero as
\be
z_{\rm B}\Big|_{T\to\infty} = N\left(\frac{\hbar\omega}{T}\right)^3
\ee
Coefficient functions $A(z_{\rm B},T)$ and $B(z_{\rm B},T)$ in this limit are 
\be
A(z_{\rm B},T) = -\frac{1}{16}z_{\rm B}, \qquad
B(z_{\rm B},T) = 4
\ee
and for $X(z_{\rm B},T)$ and $Y(z_{\rm B},T)$ one has
\be
X(z_{\rm B},T), Y(z_{\rm B},T)\to {\rm const}.
\ee
From Eq.~(\ref{eq:aA+bB=XY}) it is thus clear that
\be
-\frac{1}{16}a N\left(\frac{\hbar\omega}{T}\right)^3 + 4b \propto
\frac{1}{T^2},
\ee
which means the high-temperature limiting behavior of the parameters $a$ and $b$ as follows:
\be
a\propto T^{\nu}\quad\textrm{with}\ \nu\leq 0,
\qquad
b\propto \frac{1}{T^2}
\ee
and thus classical results are expected without any influence of the statistics deformation as $T\to\infty$.

\section{Critical temperature in the 3D case}\label{sec:Tc}
An equation to complement (\ref{eq:aA+bB=XY}) can be found, for instance, from the definition of the critical temperature of the finite weakly-interacting Bose-system.

In the thermodynamic limit, the critical temperature $T_c$ of the WNAPS system corresponding to a BEC-like transition is defined by the condition, which in the 3D case reads:
\be\label{eq:Tc-def}
N = \left(\frac{T_c}{\hbar\omega}\right)^3 
\int_0^\infty \frac{\xi^2/2}{(1+a)e_{1-b}^\xi - (1+a)}\,d\xi,
\ee
where the critical value of the fugacity is given by $z_c^{-1}=1+a$.
With linear corrections only, Eq.~(\ref{eq:Tc-def}) becomes:
\be
N =  \left(\frac{T_c}{\hbar\omega}\right)^3 \zeta(3)
\left[1-a+b\,\frac{6\zeta(4)}{\zeta(3)}\right],
\ee
where $\zeta(s)$ is the Riemann zeta-function, $\zeta(s)=\Li_s 1$.
The critical temperature of the reference Bose-system is
\be
T_c^{\rm B} = \hbar\omega \left(\frac{N}{\zeta(3)}\right)^{1/3} 
\ee
and for $T_c$ one easily obtains:
\be\label{eq:Tc=ab}
T_c = T_c^{\rm B} 
\left[1+\frac{a}{3} - b \,\frac{2\zeta(4)}{\zeta(3)}\right].
\ee

The shift of the critical temperature in a finite Bose-system of $N$ particles is given by \cite{Giorgini_etal:1996,Li_etal:1999}
\be\label{eq:DeltaTcfin}
\frac{\Delta T_c^{\rm fin}}{T_c^{\rm B}} = 
-\frac{1}{2}\frac{\zeta(2)}{[\zeta(3)]^{2/3}} N^{-1/3}
\ee
and the shift due to interaction effects is \cite{Giorgini_etal:1996}
\be\label{eq:DeltaTcint}
\frac{\Delta T_c^{\rm int}}{T_c^{\rm B}} = 
-1.33 \frac{a_s}{a_{\rm ho}} N^{1/6},
\ee
where, as above, the harmonic oscillator length $a_{\rm ho} = \sqrt{\hbar/m\omega}$ and $a_s$ is the $s$-wave scattering length. Note that in the thermodynamic limit 
\be
\frac{a_s}{a_{\rm ho}} N^{1/6} \propto \left(\omega N^{1/3}\right)^{1/2} 
= {\rm const}
\ee
does not depend on the number of particles $N$.

Comparing Eqs.~(\ref{eq:Tc=ab})--(\ref{eq:DeltaTcint}), the relation linking $a,b$  with the system parameters is obtained:
\be\label{eq:ab=system}
\frac{a}{3} - b \,\frac{2\zeta(4)}{\zeta(3)} = 
-\frac{1}{2}\frac{\zeta(2)}{[\zeta(3)]^{2/3}} N^{-1/3}
-1.33 \frac{a_s}{a_{\rm ho}} N^{1/6}.
\ee

For a system of 5000 Rb-87 atoms \cite{Giorgini_etal:1996} the ratio $a_s/a_{\rm ho}\simeq 2.6\cdot 10^{-3}$. Assuming that the $a$ parameter is entirely due to the finite-size correction and that the $b$ parameter is entirely due to interactions, the following numbers are obtained from Eq.~(\ref{eq:ab=system}):
\be
a = -0.13,\qquad
b = 0.022.
\ee

On the other hand, if Eqs.~(\ref{eq:aA+bB=XY}) and (\ref{eq:ab=system}) are solved simultaneously, the above numbers change slightly:
\be\label{eq:ab=}
a = -0.16,\qquad
b = 0.0027.
\ee

\begin{figure}[h]
\centerline{\includegraphics[scale=0.650]{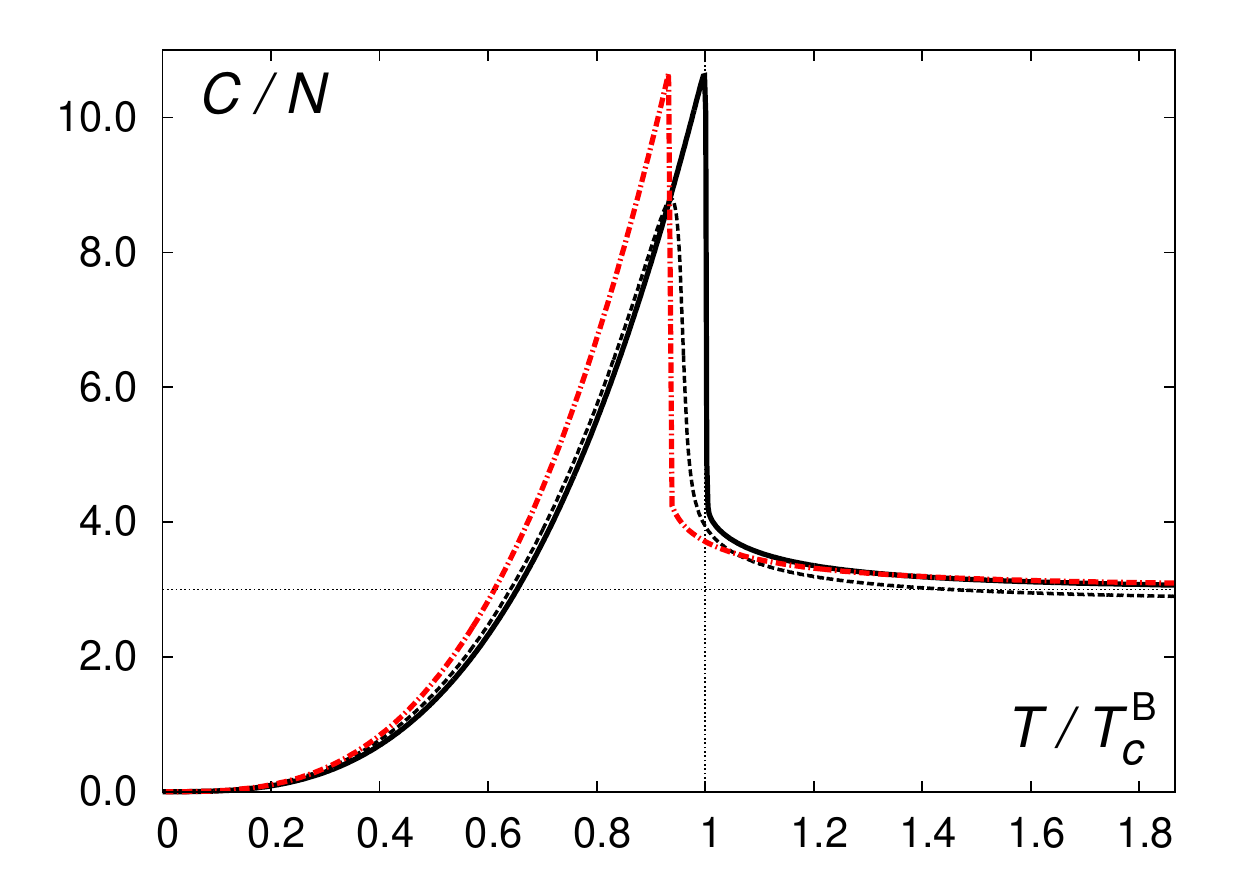}}
\caption{(Color online). Specific heat of the ideal Bose-system of 3D harmonic oscillators (black solid line --- thermodynamic limit, black dotted line --- $N=5000$) compared to the WNAPS system (red dashed-dotted line) in the thermodynamic limit with parameters given by (\ref{eq:ab=}). The discontinuity of the $C/N$ curves is observed in the thermodynamic limit at the critical temperatures but the continuous lines are drawn for better visualization.}
\label{fig:C5000}
\end{figure}

The results of calculations of the specific heat $C/N$ for a system with parameters corresponding to the above values are shown in Fig.~\ref{fig:C5000}.

Note that a smooth behavior of the specific heat in the vicinity of the critical temperature for a finite Bose-system cannot be modeled correctly by the proposed model. A possible solution is to consider a finite WNAPS system as well, which would ensure such a dependence.

\begin{figure}[h]
\centerline{\includegraphics[scale=0.650]{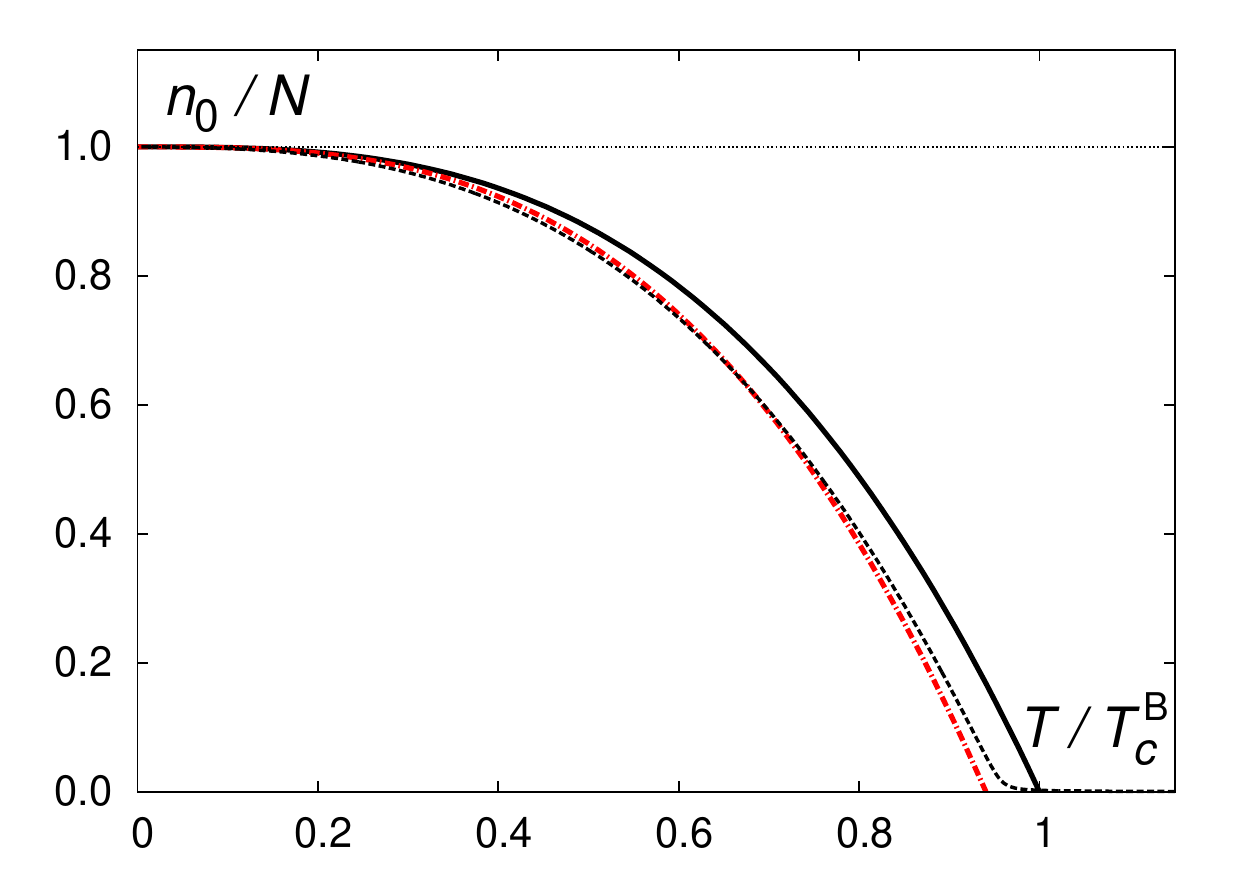}}
\caption{(Color online). Condensate fraction $n_0/N$  the ideal Bose-system of 3D harmonic oscillators (black solid line --- thermodynamic limit, black dotted line --- $N=5000$) compared to the WNAPS system (red dashed-dotted line) in the thermodynamic limit with parameters given by (\ref{eq:ab=}).}
\label{fig:n0}
\end{figure}

The fraction of particles in the ground state $n_0$, which for the WNAPS system is an analog of Bose-condensate, can be calculated quite easily as
\be
\frac{n_0}{N} = 1 - \frac{1}{N}
\frac{1}{1+a}\left(\frac{T}{\hbar\omega}\right)^3 
\int_0^\infty \frac{\xi^2/2}{e_{1-b}^\xi - 1}\,d\xi
\ee
Using the parameters from (\ref{eq:ab=}), the following temperature dependence is obtained for $N=5000$:
\be
\frac{n_0}{N} = 1 - \frac{1.45}{N}
\left(\frac{T}{\hbar\omega}\right)^3.
\ee
The comparison of the above result with reference Bose-systems is shown in Fig.~\ref{fig:n0}. The shapes of these dependences are very similar to those reported in \cite{Giorgini_etal:1996,Ensher_etal:1996}.

\section{Specific heat of model WNAPS systems}\label{sec:CN}
Having estimated the values of statistics parameters $a$ and $b$, it is possible to present some results illustrating the behavior of thermodynamic functions (namely, the specific heat) of model systems obeying the weakly nonadditive Polychronakos statistics. Below, the calculations are made for two modifications of the statistics. In the first one, the parameters $a,b$ are kept constant with respect to temperature. In the second one, the parameter $a$ remains constant but the parameter $b=2\eta T/\hbar\omega$, so that the summands with $b$ in Eq.~(\ref{eq:nj-WNPS}) and with $\Delta\eps_j$ in Eq.~(\ref{eq:nj-Bose}) become similar.

Indeed, as we assume the second of the abovementioned statistics modifications, the spectrum shift is
\be
\Delta\eps_j = -\frac{b\eps_j^2}{2T} = 
-\eta\frac{\eps_j^2}{\hbar\omega} = -\eta\hbar\omega j^2.
\ee
Curiously, such a dependence of the excitation spectrum appears in the problems within deformed Heisenberg algebras. Namely, for the harmonic oscillator with the commutation relation for the coordinate and momentum operators given by
\be
[\hat x,\hat p]=i\hbar(1+\beta \hat p^2)
\ee
the spectrum is \cite{Kempf:1995,Quesne&Tkachuk:2003}
\be
\eps_j = \hbar\bar\omega j + \frac{\beta}{2} j^2,
\ee
where $\bar\omega$ denotes some constant. Due to an extremely small estimated values of $\beta$, however, its effect on the thermodynamic properties is unobservable.

\begin{figure}[h]
\centerline{\includegraphics[scale=0.80]{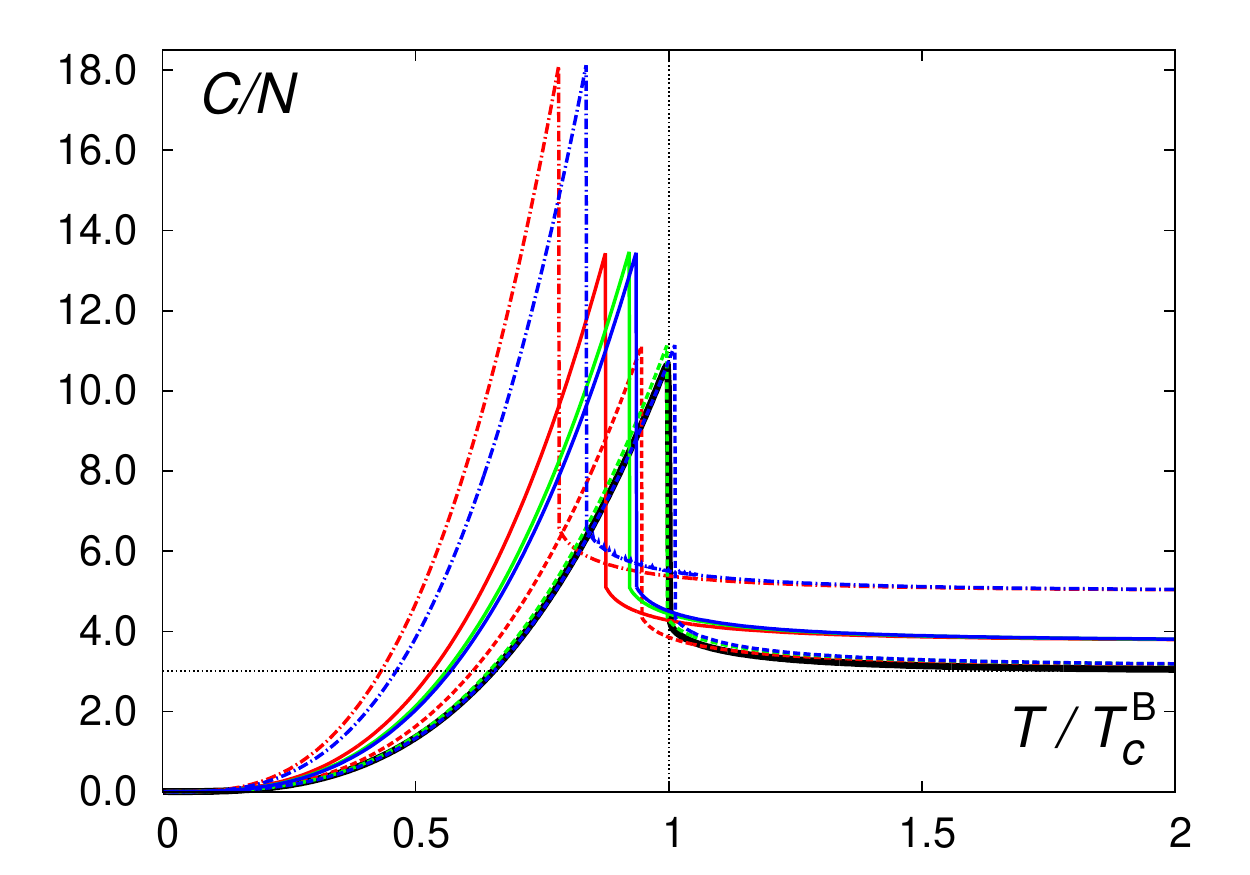}}
\caption{(Color online). Specific heat of a model WNAPS system with $b={\rm const}$ in the thermodynamic limit for different values of the statistics parameters. 
Dashed-dotted lines --- $b=0.1$; solid lines --- $b=0.05$; dashed lines --- $b=0.01$. Color correspondence: red --- $a=-0.1$, green --- $a=+0.05$, blue --- $a=+0.1$. Black solid line corresponds to the reference Bose-system.}
\label{fig:C1000_1}
\end{figure}

\begin{figure}[h]
\centerline{\includegraphics[scale=0.80]{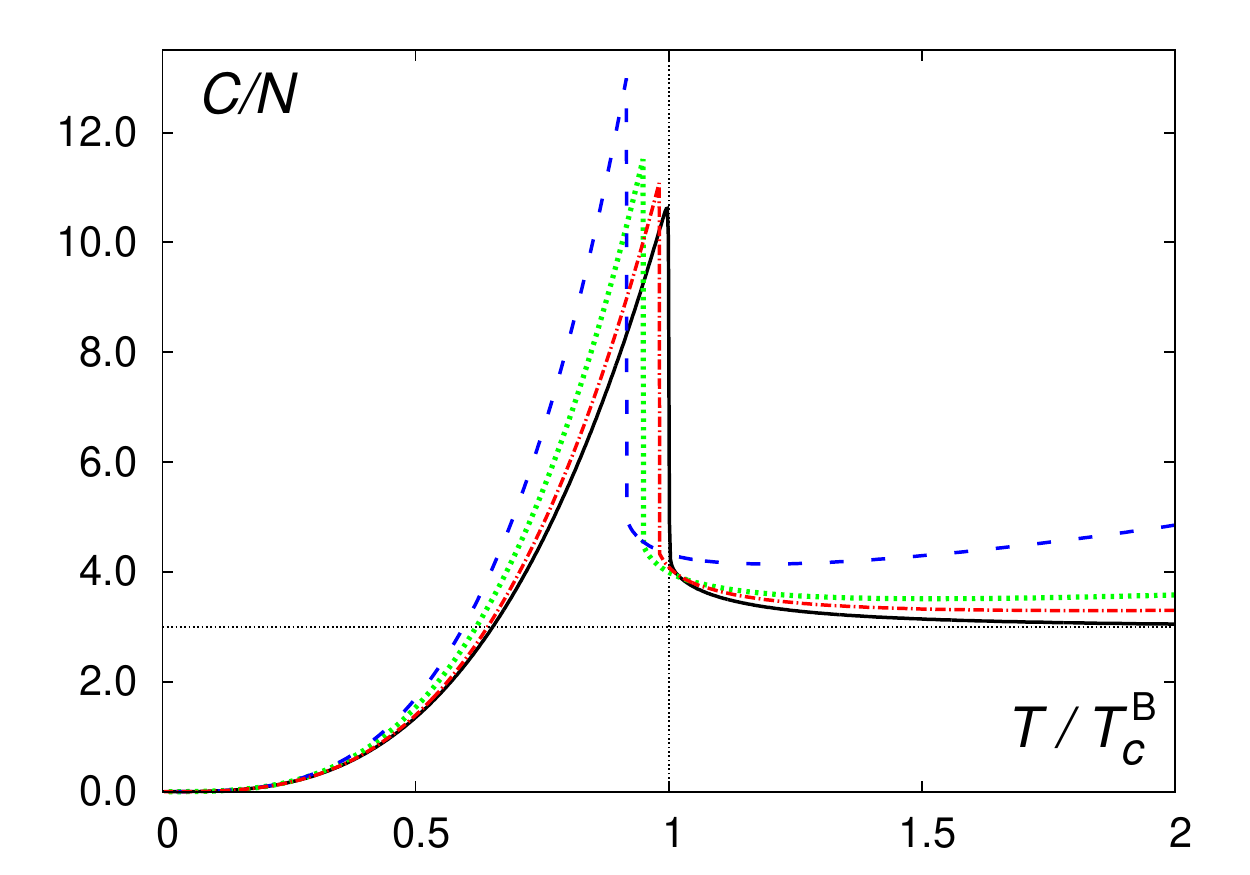}}
\caption{(Color online). Specific heat of a model WNAPS system with $b=2\eta T/\hbar\omega$ in the thermodynamic limit for different values of the statistics parameters. 
Dashed-dotted line, red --- $a=0,\eta=0.0005$; dotted line, green --- $a=-0.05,\eta=0.001$; dashed line, blue --- $a=-0.01,\eta=0.0025$. Black solid line corresponds to the reference Bose-system.}
\label{fig:C1000_2}
\end{figure}

The results of calculations of the specific heat are shown in Figs.~\ref{fig:C1000_1}--\ref{fig:C1000_2} in comparison to a reference Bose-system. The choice of the values of the statistics parameters is made according to the estimations from the previous Section. Note that $b\propto T$ model is valid only in a limited temperature domain since the system ceases to be weakly nonadditive as $T\to\infty$.

In the model with $b={\rm const}$, the asymptotic value of the specific heat at $T\to\infty$ can be estimated as follows. The fugacity tends to zero as $T\to\infty$, so Eq.~(\ref{eq:N=int_Bose}) in the case of harmonic oscillator problem under consideration simplifies to
\be
N = \left(\frac{T}{\hbar\omega}\right)^3 \frac{z}{2} 
\int_0^\infty d\xi\, \frac{\xi^2}{e_{1-b}^{\xi}}.
\ee
Applying the relation \cite{Yamano:2002}
\be
\left[e^{f(x)}_q\right]^p = e^{pf(x)}_{1-(1-q)/p},
\ee
this integral can be calculated using
\be
\int_0^\infty d\xi\, \xi^{k-1} e_{1+b}^{-\xi} = 
\frac{b^{-k}\Gamma(1/b-k)\Gamma(k)}{\Gamma(1/b)}.
\ee
The energy equals in the same limit to
\be
E = \hbar\omega\left(\frac{T}{\hbar\omega}\right)^4 \frac{z}{2} 
\int_0^\infty d\xi\, \frac{\xi^3}{e_{1-b}^{\xi}}.
\ee
After simple transformations, one obtains for $z$
\be
z = N\left(\frac{\hbar\omega}{T}\right)^3 
\frac{(1 - 6 b + 11 b^2 - 6 b^3)}{2}
\ee
and the heat capacity is given by 
\be
\left.\frac{C}{N}\right|_{T\to\infty} = \frac{3}{1-4b}
\ee
showing no dependence on the value of the $a$ parameter.

\section{Conclusions}\label{sec:conclusions}
In summary, a two-parametric modification of statistics was proposed, which can be used in particular to model a weakly-interacting Bose-system. It was shown that the parameters of the introduced weakly nonadditive Poly\-chro\-nakos statistics can be linked to effects of inter\-actions as well as finite-size corrections.

A simplified WNAPS model was used to describe the system of 5000 harmonically trapped Rb-87 atoms. The calculations of the specific heat $C/N$ of the 3D isotropic harmonic oscillators were also made for several sets of values of the statistics parameters $a,b$ to demonstrate the temperature behavior of $C/N$ in the domain including the BEC-like phase transition point.

It is expected that WNAPS can provide an alternative mathematical model for Bose-systems with weak interatomic interactions and/or finite number of particles. Its application to correctly reproduce the critical behavior in the vicinity of the transition point requires additional tests based on experimental observations. With minor modifications, the model can be employed for other related systems, in particular lower-dimensional oscillators corresponding to trapped bosons.

Some other physical objects, for which the proposed two-parametric statistics can be used, include anyons and particles in spaces with minimal length. While the latter correspond to the deformed Heisenberg algebra discussed in Sec.~\ref{sec:CN}, the application to the anyonic statistics is elucidated by the possibility to establish an approximate correspondence with the nonadditive Polychronakos statistics from expressions for virial coefficients. An effective description of long-range interactions and other complex behavior can be expected from this statistics and these issues are the subjects of further studies.

\section*{Acknowledgments}
I am grateful to Referees for helpful comments an suggestions.

This work was partly supported by Project $\Phi\Phi$-110$\Phi$ (registration number 0112U001275) from the Ministry of Education and Sciences of Ukraine.

\end{document}